\documentclass[leter,twocolumn,11pt]{article}
\usepackage[spanish]{babel}
\usepackage{amssymb}
\usepackage{ae}
\usepackage[all]{xy}
\usepackage{tikz}
\usepackage{graphicx}
\usetikzlibrary{backgrounds,calc}
\usepackage{amsmath}
\usepackage[]{cite}   
\usepackage{anysize}  
\marginsize{1.0cm}{1.0cm}{1.5cm}{2.5cm}
\begin{document}
\renewcommand{\tablename}{Tabla}
\renewcommand{\figurename}{Fig.}
\renewcommand{\abstractname}{}
\renewcommand{\thefootnote}{\arabic{footnote}}

\title{
        {\LARGE Tautochrone in the damped cycloidal pendulum\\
        \footnotesize   \textit{(Tautochrone no pêndulo cicloidal amortecida)}}
}
\author
{
Paco H. Talero L.$^{1}$,  César A. Herreño-Fierro$^{2}$ and Fernanda Santana G. $^{3}$ \\  
$^{1}$ \small Grupo MatCIng, Dpt. de Ciencias Naturales, Universidad Central, Cl. 21 $\#$4-40, Bogotá D.C. Colombia\\
$^{2}$ \small Facultad de Ciencias y Educación, Universidad Distrital Francisco José de Caldas, Cra. 7 40B-53, Bogotá-Colombia\\
$^{3}$ \small Dpt. de Ciencias Naturales, Universidad Central, Cl. 21 $\#$4-40, Bogotá D.C. Colombia} 
\date{}

\twocolumn
[
\begin{@twocolumnfalse}

\maketitle 
\begin{abstract}
The tautochrone on a cycloid curve is usually considered without drag force. In this work, we investigate the motion of a damped cycloidal pendulum under presence of 
a drag force. Using the Lagrange formulation, and considering linear dependence with velocity for damping force, we found the dynamics of the system to remain tautochrone. This 
dictates the possibility for studying the tautochrone experimentally, e.g. the cycloidal pendulum in water or oil. 

A tautócrona em uma curva cicloide geralmente ocorre sem força de arrasto, o objetivo deste estudo foi pesquisar se a tautócrona em
um pêndulo cicloidal se mantem na presença de uma força de arrasto linear, com a velocidade. Para isto foi utilizada a formulação
de Lagrange. Constatamos que a tautócrona se mantem, mas apenas no caso em que o pêndulo cicloidal se comporta como um oscilador harmónico
amortecido. Os resultados deste trabalho oferecem possibilidades de observação da tautócrona em sala de aula com laboratório,
por exemplo, o pêndulo cicloidal em água ou óleo.

\textbf{Palavras-chave:} tautócrona, pêndulo cicloidal, oscilador harmônico amortecido, força de arrasto.\\
\end{abstract}
\end{@twocolumnfalse}
]

\section{Introduction}

The cycloid has received much attention due to its mathematical and physical properties, which offers many pedagogical benefit. There are two well known physical properties: the brachistochrone and the tautochrone \cite{villa,Figue,MASH}. The brachistochrone is the curve upon which the motion of a particle between two points in a vertical plane lasts the shortest time, i.e. $A\rightarrow O$ in Fig.\ref{zen}. The tautochrone is the curve upon which it takes a particle the same time to go from any point to the lowest point on the curve, e.g. $A\rightarrow O$ or $B \rightarrow O$ in Fig.\ref{zen}.\\

It is generally accepted that the tautochrone only occurs for the frictionless motion of a particle through a cycloid track or a cycloidal pendulum. In this paper we proof that this is false. In the following section we discuss the tautochrone in a cycloidal pendulum with drag force, trough a simple Lagrangian analysis.
\begin{figure}[!htp]
\begin{center}
	\begin{tikzpicture}[scale=0.8]
	\draw[->,black,line width=1.0pt] (-4.5cm,0cm)--(4.5cm,0cm);
	\draw[->,black,line width=1.0pt] (0.0cm,-0.1cm)--(0cm,4.5cm);
	\draw[->,black,line width=1.5pt] (-4.0cm,1.5cm)--(-4.0cm,0.75cm);
	\draw [dashed,gray,line width=1.5pt,domain=0.0:2.0*pi] plot ({\x-3.1415-sin(\x r)},{1.0+cos(\x r)});
	\draw [thick, fill=gray!20,domain=0.0:1.0*pi] plot ({\x-sin(\x r)},{3.0+cos(\x r)});
	\draw [thick, fill=gray!20,domain=0.0:-1.0*pi] plot ({\x-sin(\x r)},{3.0+cos(\x r)});
	\draw[dashed,black,line width=0.5pt] (3.1415cm,0cm)--(3.1415cm,2.0cm);
	\draw[dashed,black,line width=0.5pt] (-3.1415cm,0cm)--(-3.1415cm,2.0cm);
	\draw[dashed,black,line width=0.5pt] (0.0cm,1.0cm)--(2.65cm,1.0cm);
	\draw[black,line width=0.5pt] (-2.57cm,1.0cm)--(-0.7cm,2.87cm);
	\draw[black,line width=0.5pt] (2.57cm,1.0cm)--(0.7cm,2.87cm);
	\fill[black](-2.57cm,1.0cm) circle (0.12cm); 
	\fill[black](2.57cm,1.0cm) circle (0.12cm); 
	\fill[black](-3.1415cm,2.0cm) circle (0.12cm); 
	\coordinate [label=below:\textcolor{black} {$y$}] (x) at  (0.1cm,5.0cm);
	\coordinate [label=below:\textcolor{black} {$x$}] (x) at  (4.7cm,0.2cm);
	\coordinate [label=below:\textcolor{black} {$O$}] (x) at  (0.0cm,-0.05cm);
	\coordinate [label=below:\textcolor{black} {$\pi R$}] (x) at  (3.1415cm,-0.05cm);
	\coordinate [label=below:\textcolor{black} {$-\pi R$}] (x) at  (-3.2cm,-0.05cm);
	\coordinate [label=below:\textcolor{black} {$4R$}] (x) at  (0.5cm,4.2cm);
	\coordinate [label=below:\textcolor{black} {$2R$}] (x) at  (3.5cm,2.3cm);
	\coordinate [label=below:\textcolor{black} {$A$}] (x) at  (-3.4cm,2.3cm);
	\coordinate [label=below:\textcolor{black} {$B$}] (x) at  (-2.65cm,1.0cm);
	\coordinate [label=below:\textcolor{black} {$C$}] (x) at  (2.65cm,1.0cm);
	\coordinate [label=below:\textcolor{black} {$y$}] (x) at  (-0.18cm,1.3cm);
	\coordinate [label=below:\textcolor{black} {$\overrightarrow{g}$}] (x) at  (-3.7cm,1.5cm);
	\end{tikzpicture}
\caption{Cycloidal pendulum: a particle with mass $m$ tied to a string of length $4R$ oscillate on a cycloid with
         generator circle of ratio $R$.}
\label{zen}
\end{center}
\end{figure}
\section{Motion under damping force is also tautochrone}
Consider the cycloidal pendulum shown in Fig.\ref{zen}. In this inertial frame the particle
has a motion through the cycloid, so that its parametric equations are
\begin{equation}\label{cyx}
x=R\left( \theta-\pi-\sin\theta  \right)
\end{equation}
and
\begin{equation}\label{cyy}
y=R\left(1+\cos\theta \right)
\end{equation}
with $0 \leq \theta \leq 2\pi$. The arc length is $s=-4R\cos \frac{\theta}{2}$, thus $s>0$ for $\theta>\pi$
and $s<0$ for $\theta<\pi$.\\

From Fig \ref{zen}, the potential energy of particle is $E_p=mgy$.  According to
Eq.(\ref{cyy}) it also can be written as
\begin{equation}\label{cyy}
E_p=\frac{1}{2} k s^{2}
\end{equation}
where $k=\frac{mg}{4R}$.\\

The  Lagrangian of system is
\begin{equation}\label{Lg}
L=\frac{1}{2}m \dot{s}^{2}-\frac{1}{2}ks^{2}
\end{equation}
and the Lagrange's motion equation with drag force $\overrightarrow{f}=-b\dot{s} \widehat{u}_{T}$ is
\begin{equation}\label{Lg2}
\frac{d}{dt} \left( \frac{\partial L}{\partial \dot{s}} \right)-\frac{\partial L}{\partial s}=-b\dot{s}
\end{equation}
where the generalized coordinate is the arc length $s$ and $\widehat{u}_{T}$ is a unit vector
tangent to the curve.\\

Let us define $\omega_{o}^{2}=\frac{g}{4R}$ and $\gamma=\frac{b}{2m}$, thus we can write the solutions of Eq.(\ref{Lg2}) as
\begin{equation}\label{ecd}
\ddot{s}+2\gamma \dot{s}+\omega_{o}^{2}s=0,
\end{equation}
it is the  differential equation of damped harmonic oscillator. We are interested in the case $\omega_{o}>\gamma$. Therefore, the solution of  Eq.(\ref{ecd}) for $v_o=0$ is
\begin{equation}\label{sol}
s=s_oe^{-\gamma t}\cos \left(\omega t \right)
\end{equation}
where $\omega^{2}=\omega_{o}^{2}-\gamma^{2}$, $T=\frac{2\pi}{\omega}$, and $s_o$ is the initial position
of the particle.\\
\begin{figure}[!htp]
\begin{center}
	\begin{tikzpicture}[domain=0:1.57,scale=2.5]
	\draw [->,black,line width=1.5pt] (-0.05cm,0cm) -- (1.7cm,0cm);
	\draw [->,black,line width=1.5pt] (0.0cm,-1.2cm) -- (0.0cm,1.7cm);
	\coordinate [label=below:\textcolor{black} {$t$}] (x) at  (1.74cm,0.12cm);
	\coordinate [label=below:\textcolor{black} {$s(t)$}] (x) at  (0.0cm,1.95cm);
     \draw[color=black,line width=1pt]   plot (\x,{1.5*exp(-1.0*\x)*cos(1.0*\x r)});
     \draw[color=black,line width=1pt]   plot (\x,{0.5*exp(-1.0*\x)*cos(1.0*\x r)});
     \draw[color=black,line width=1pt]   plot (\x,{-1.0*exp(-1.0*\x)*cos(1.0*\x r)});
     \coordinate [label=below:\textcolor{black} {$\frac{T}{4}$}] (x) at  (1.57cm,0cm);
     \coordinate [label=below:\textcolor{black} {$O$}] (x) at  (-0.1cm,0cm);
     \coordinate [label=below:\textcolor{black} {$s_{o}$}] (x) at  (-0.11cm,1.6cm);
     \coordinate [label=below:\textcolor{black} {$s_{1}$}] (x) at  (-0.11cm,0.6cm);
     \coordinate [label=below:\textcolor{black} {$s_{2}$}] (x) at  (-0.11cm,-0.9cm);

	\end{tikzpicture}
\caption{Position vs time of the la particle in movement on cycloidal from $t=0$ to $t=\frac{T}{4}$.}
\label{Temp}
\end{center}
\end{figure}
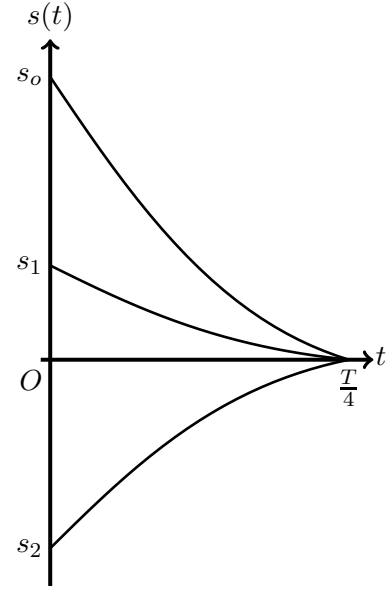

Fig.\ref{Temp} shows the plot of Eq.(\ref{sol}) for three different initials  positions $s_o$, $s_1$ and $s_2$. It is clear that the time it takes to the particle to arrive the $O$ position does no dependent on the  initial position, showing the tautochrone even in presence of a damping force.

\section{Conclusions}
In this work we reported the tautochrone in the cycloidal pendulum in presence of a damping force using a simple Lagrangian analysis. These results could be used for the experimental physics class or as a simple experimental demonstration of the tautochrone with a cycloidal pendulum  into air water, oil or other fluid.

\section*{Acknowledgments}
 Authors would like to thank the Faculty of Engineering of Universidad Central for offering of the course of Lagrangian mechanics, and Professor Olga Lucía Castiblanco from Universidad Distrital Francisco José de Caldas for her enriching discussions.
\renewcommand{\refname}{Bibliography}


\begin{thebibliography}{99}
\bibitem{villa} J. Z. Villanueva, Am. J. Phys. \textbf{53}, 490-491 (1985).
\bibitem{Figue} D. Figueroa, G. Gutierrez, and C. Fehr, Phys. Teach. \textbf{35}, 494-498 (1997).
\bibitem{MASH} M. Sawicki, Phys. Teach. \textbf{43}, 236-238 (2005).

\end{thebibliography}
\end{document}